# Automatic segmentation of CT images for ventral body composition analysis

Yabo Fu, Joseph E. Ippolito, Daniel R. Ludwig, Rehan Nizamuddin, Harold H. Li, Deshan Yang*

Washington University School of Medicine, 660 S Euclid Ave, Campus Box 8131, St Louis, MO 63110

*Corresponding author:

Deshan Yang

Washington University in Saint Louis

yangdeshan@wustl.edu

# Abstract

**Purpose:**

Body composition is known to be associated with many diseases including diabetes, cancers and cardiovascular diseases. In this paper, we developed a fully automatic body tissue decomposition procedure to segment three major compartments that are related to body composition analysis – subcutaneous adipose tissue (SAT), visceral adipose tissue (VAT) and muscle. Three additional compartments - the ventral cavity, lung and bones were also segmented during the segmentation process to assist segmentation of the major compartments.

**Methods:**

A convolutional neural network (CNN) model with densely connected layers was developed to perform ventral cavity segmentation. An image processing workflow was developed to segment the ventral cavity in any patient's CT using the CNN model, then further segment the body tissue into multiple compartments using hysteresis thresholding followed by morphological operations. It is important to segment ventral cavity firstly to allow accurate separation of compartments with similar Hounsfield unit (HU) inside and outside the ventral cavity.

**Results:**

The ventral cavity segmentation CNN model was trained and tested with manually labelled ventral cavities in 60 CTs. Dice scores (mean ± standard deviation) for ventral cavity segmentation were 0.966±0.012. Tested on CT datasets with intravenous (IV) and oral contrast, the Dice scores were 0.96±0.02, 0.94±0.06, 0.96±0.04, 0.95±0.04 and 0.99±0.01 for bone, VAT, SAT, muscle and lung, respectively. The respective Dice scores were 0.97±0.02, 0.94±0.07, 0.93±0.06, 0.91±0.04 and 0.99±0.01 for non-contrast CT datasets.

**Conclusion:**

A body tissue decomposition procedure was developed to automatically segment multiple compartments of the ventral body. The proposed method enables fully automated quantification of 3D ventral body composition metrics from CT images.

# 1   Introduction

Body composition metrics such as body mass index (BMI), adiposity, and lean body mass are important health indicators that are used as evidence to describe the health condition of an individual or a population[1-3]. Studies of body composition can improve our understanding of the impact of body composition on common diseases including cardiovascular diseases[4], hypertension and diabetes[5] and even cancer[6]. For example, studies have shown that an increased ratio of adipose to lean body mass (i.e., sarcopenic obesity) can be linked to an increased risk of these diseases[7-9]. Adipose tissue is subdivided into subcutaneous adipose tissue (SAT) and visceral adipose tissue (VAT). SAT is located outside the ventral body cavity and VAT is located inside. The ratio between VAT and SAT was reported to be an independent predictor of mortality for cardiovascular events and cancer[6,10,11]. In addition to adiposity, muscle mass and bone mineral density are important indicators in cancer prognosis and disease mortality[12-15]. Thus, the ability to efficiently and routinely calculate body composition metrics in the clinical setting represents a significant unmet need that can allow clinicians to risk-stratify patients. However, body composition metrics are not routinely calculated in the clinic due to the need of laborious segmentation of multiple body compartments and the lack of automation. In practice, manual or semi-automatic segmentation of SAT, VAT and muscle on a single CT slice are routinely used to approximate the 3D body composition[16,17]. However, this approach has several limitations including 1) the selected 2D slice may not accurately represent the total 3D ventral body composition[18], 2) manual contouring may be laborious and time consuming even on a single slice, and 3) manual contouring is subject to inter-observer variability.

Several methods have been published to automate the segmentation process for body composition analysis from CT images[13,19-23]. Lee *et al.* used an active contour model to iteratively modify the external contour of the body to find the inner boundary of the SAT[21]. However, this method is not robust as it fails on slices where small sections of muscle, vessels, or other small anatomical structures appear inside the SAT and on slices where the SAT is not homogeneous due to anatomic variability and noise. Weston *et al.* proposed a U-Net like convolutional neural network (CNN) model to segment four compartments including SAT, VAT, bone and muscle[19]. This method was able to automatically quantify body composition metrics in 3D CT images. They only focused on the abdomen ranging from the L1 to L5 vertebrae. Trained on 2D transaxial CT images at the level of the L3 vertebra, their model generalized well at other vertebral levels such as L4. Wang *et al.* used CNN to segment abdominal adipose tissue including SAT and VAT[20]. Lee *et al.* used a CNN[13] to segment abdominal muscle. To summarize, none of these methods were designed to segment the whole ventral body compartment including four major compartments related to body composition: SAT, VAT, muscle and bone.

In this study, we present a fully automated procedure to accurately segment these ventral body compartments for routine body composition measurements in the clinic. Our CNN-based approach for ventral cavity segmentation is distinctly advantageous in that it enables automatic segmentation of multiple body compartments using a simple HU thresholding process. Important benefits of using the simple HU thresholding step over the learning-based tissue classification methods are that it is computationally efficient, does not rely on the availability and accuracy of manual annotations for model training, and is robust with both contrast-enhanced and non-contrast CT datasets. Additionally, our method provides multiple body compartment segmentations of the entire ventral cavity (i.e., thoracic, abdominal, and pelvic cavity), which is important for accurate and comprehensive body composition index calculations.

## 2 Materials and methods
### 2.1 Image datasets

A total of 60 CT datasets were used in this study. 30 CTs with oral and/or intravenous (IV) contrast were obtained from institutional archives and were from individuals without evidence of malignancy in the ventral cavity or surgical manipulation. The other 30 CT datasets, 15 with IV contrast and 15 without IV contrast, were from The Cancer Imaging Archive (TCIA) with collection IDs of 'TCGA-KIRC[24]' and 'QIN-HEADNECK'[12]. The original slice thicknesses ranged from 2 mm to 5 mm. All datasets were resampled to have a consistent slice thickness of 2 mm. The 30 institutional datasets were manually segmented by trained radiologists to generate ventral cavity contour ground truth and were used to train and test the ventral cavity segmentation CNN model. The 30 TCIA datasets were used as independent validation datasets to evaluate the whole workflow. For each of these 30 TCIA cases, 5 transaxial slices that cover the entire ventral cavity were uniformly sampled and manually contoured to evaluate the segmentation accuracy of the proposed method.

### 2.2 The two-step automatic-segmentation workflow

Though lung, bone, adipose tissue and muscle have discrete Hounsfield unit (HU) ranges, it is very challenging for many reasons to rely solely on HU thresholding techniques to segment them. Anatomic structures inside the ventral cavity including organs, bowel and vessels have overlapping HU with these body compartments. Adipose tissue such as SAT and VAT have the same HU, which makes it impossible to separate them without knowing the ventral cavity boundary. In CT images acquired with both oral and IV contrast, bowel and vessels have HU similar to bone, increasing the difficulty for tissue classification. Noise, artifacts, and partial volume effects further increase the segmentation difficulty of the CT datasets.

We overcame these challenges by segmenting the ventral cavity prior to body compartment segmentation in a two-step procedure, shown in Fig. 1. First, a CNN model was applied to segment the ventral cavity. Second, body compartments were segmented using HU thresholding and morphological smoothing with the ventral cavity that allowed for the separation of VAT and internal organs from SAT and skeletal muscle.

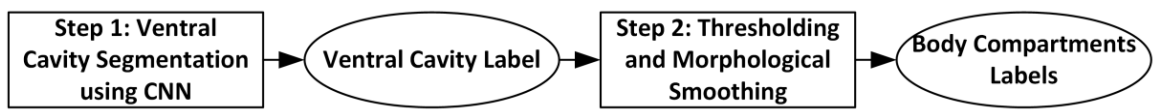

Fig. 1: The workflow to automatically segment the ventral body cavity and multiple body compartments.

### 2.3 Step 1 - ventral cavity segmentation using a CNN model

The ventral cavity includes the thoracic cavity, abdominal cavity and pelvic cavity. With the segmented ventral cavity, the muscle outside the cavity can be separated from the organs with similar HU inside the cavity, and SAT outside the cavity can be easily separated from VAT inside the cavity.

#### 2.3.1 CNN model design, implementation and training

We developed a CNN model to segment the ventral cavity. The architecture of the CNN model is shown in Fig. 2. The network consists of 12 convolutional layers, one max-pooling layer and three deconvolutional layers. The input image patches of size 512×512×3, representing three consecutive slices, were uniformly sampled from original images prior to feeding the CNN. To increase the size of the receptive field and save memory, the input images were first processed using a max pooling layer and two convolutional layers with strides of 2×2×1. Input image feature sizes were reduced from 512×512×3 to 64×64×3. A dense block consisting of 10 densely connected convolutional layers (Conv 3 to Conv 12) was used after the second convolutional layer. We chose the dense block design because it was efficient in high-level feature learning and could potentially alleviate the problem of gradient exploding/vanishing by encouraging information propagation from previous layers[25]. Skip connections were used to concatenate the feature maps from the encoding path and the decoding path. Three consecutive transposed convolution layers were used to up-sample the feature maps to the size of the original input images. The batch normalization layers (BN) before each convolution layer were designed to reduce internal covariate shift of the input data to each layer[26]. To alleviate the problem of potential overfitting, a dropout layer was appended to each Conv layer with an empirical dropout rate of 0.2[27].

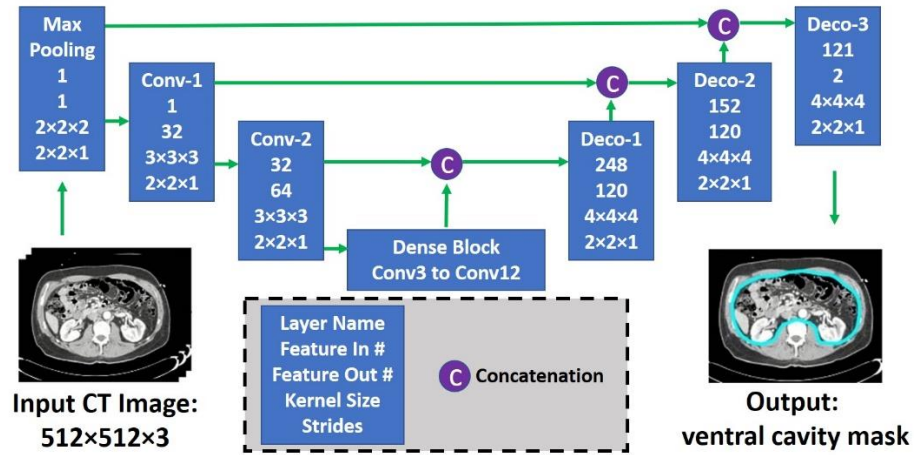

Fig. 2: The architecture of the proposed CNN model for ventral cavity segmentation.

### 2.3.2 Training data preparation - manual ventral cavity segmentation

To generate high quality contours of the ventral cavity for network training, a series of rules were defined for the manual segmentation. These rules, listed below, were important to minimize inter-observer variations especially for the pelvic region where the definition of pelvic cavity may be ambiguous.

a. Exclude all muscles associated with the pelvis (i.e. obturator, iliacus, psoas, iliopsoas, piriformis).

b. Begin caudal limit of segmentation at the inferior surface of the pubic symphysis and proceed cranially.

c. Segment the peritoneal cavity caudally using the puborectalis muscle posteriorly and laterally, obturator muscles anteriorly and laterally, and pubic bone anteriorly.

d. If the ischiococcygeus muscle appears separate from the puborectalis, use that muscle for the contouring that will ultimately become confluent with the puborectalis.

All segmentations were done by two radiology residents and one medical physics resident, each with 2 to 4 years of experience for interpreting CT images and supervised by a fellowship trained abdominal radiologist with 7 years of post-fellowship experience. Examples of the manual segmentation are shown in Fig. 3.

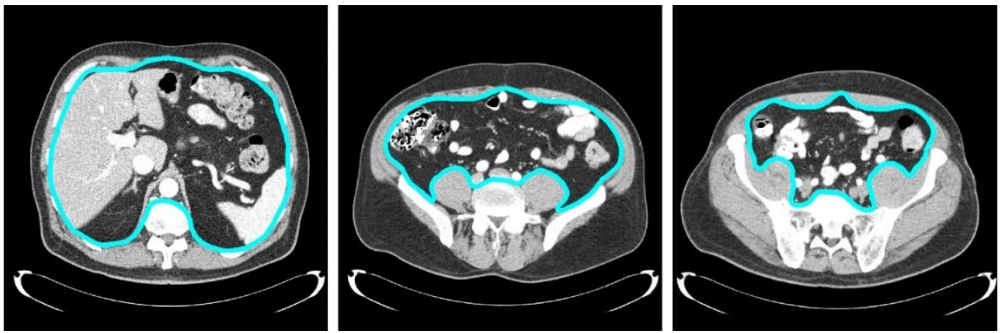

Fig. 3: Manual segmentation of the thoracic cavity, abdominal cavity and pelvic cavity.

### 2.3.3 Model implementation, training and testing

The network was implemented using Tensorflow in Python. The training parameters were initialized randomly using a Gaussian distribution (mean = 0, std = 0.01). The network was trained using the Adam optimizer with a constant learning rate of $1e^{-4}$. The loss function of the network was the binary cross entropy between the predicted labels and the ground truth labels. The learning process was consistently assessed by 5 validation cases every 500 iterations. The training process was stopped if the loss did not improve for 10 consecutive evaluations, i.e. 5000 iterations. A GeForce GTX 1080 Ti GPU with 11GB RAM and 3584 CUDA cores was used for training. The training process converged in 20 hours. Data augmentation of rotation, scaling and translation was performed during training to increase the trained network's generalizability and robustness.

## 2.4 Step 2 - ventral body compartment segmentation

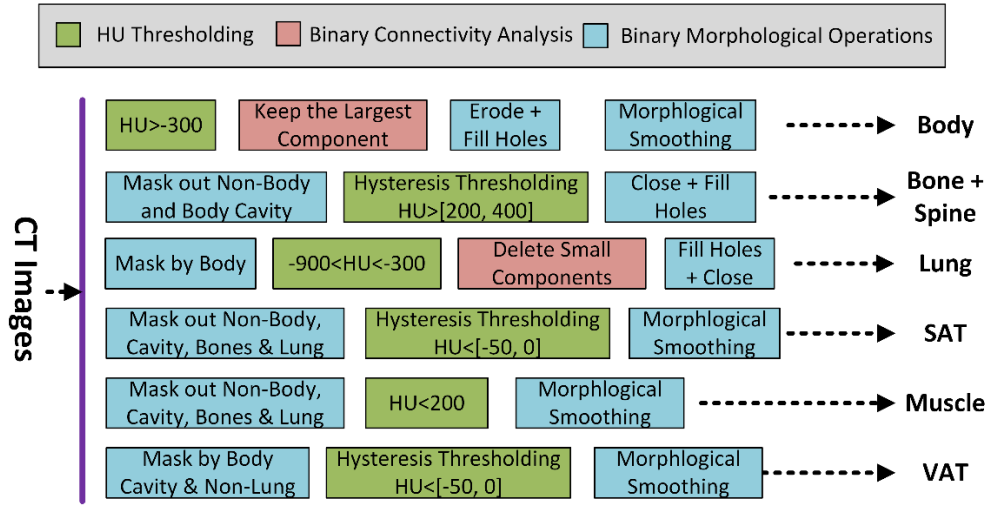

Fig. 4: Detailed procedures for HU thresholding and morphological operations in step 2.

The body compartment segmentation workflow is shown in Fig. 4. Body, bone + spine, lung, SAT, muscle, and VAT are segmented in order. Body stands for all tissues inside the skin surface. The segmentation order was designed so that the compartments which can be segmented easily and accurately, e.g. body, bone, and lung, were segmented first. The accurate segmentations of body, bony structures, and lung, and the previously segmented ventral cavity were then utilized to support the segmentation of the other more difficult structures, i.e., SAT, muscle, and VAT.

Body (i.e. skin) segmentation was performed by HU thresholding and keeping only the largest connected binary masks. The automatic segmentation of body mask often needed post-processing since the patient body was in contact with CT table and sometimes cables and wires. To have a clean body mask, image erosion was performed using a 3D spherical structuring element with radius 4 mm.

After the body was segmented, bone including the spine were segmented. Hysteresis thresholding was used to segment the bone inside the segmented body. Tissues inside the segmented ventral cavity were also excluded to prevent the GI organs imaged with oral contrast from being incorrectly segmented as bone. In hysteresis thresholding, two mask volumes were first obtained by thresholding the CT image using 200 HU and 400 HU, respectively. A flood-fill operation was then performed by growing the regions in the 400 HU mask volume into the connected voxels in the 200 HU mask volume. The hysteresis thresholding method was used based on the observations that all voxels with HU value greater than the higher threshold, i.e. 400 HU, are bone and the voxels with intensity greater than the lower threshold, i.e. 200 HU, are likely to be bone only if they were in contact with voxels greater than the higher threshold. The hysteresis thresholding alone was not enough because structures with lower HU such as bone marrow, spinal canal, inter-vertebral disks and cartilage between bones often cause the mask to have small holes and minor defects. A binary close operation with a 3D spherical structuring element of radius 16 mm and a binary fill operation were performed to clean the masks by filling small holes and removing minor defects. The same hysteresis thresholding procedure plus morphological post-processing were used to segment bone in our previous skeleton segmentation study[28] and had shown to be both robust and accurate.

Lung was segmented in three steps. First, HU thresholding was set between -900 and -300. Second, holes due to pulmonary vessels were filled using binary close operation with spherical structuring element of radius 5 m. Third, the isolated sub-volumes were identified, and any sub-volume was excluded if it was not one of the two biggest sub-volumes (i.e., the left and right lungs) or its volume was less than 200 cc. This third step was designed to prevent misclassification of air pockets in the GI organs as lung.

SAT was segmented by hysteresis thresholding <0 HU. Of note, there are no universal standards for HU thresholding of adipose tissue that varies considerably among studies[2,29]. The threshold <0 HU (inside the body but outside the ventral cavity) was found to work better for both contrast and non-contrast CTs in this study. Hysteresis thresholding was applied, instead of a simple threshold, to improve the robustness. Specifically, two mask volumes were obtained by thresholding the CT image outside the segmented ventral cavity but excluding the segmented bone using -50 HU and 0 HU, respectively. A flood-fill operation was then performed by growing the regions mask volume of HU<-50 into the connected voxels in the mask volume of HU<0. The SAT mask was post-processed using binary open then close operations with a spherical structuring element of radius 1mm to remove very small labelling.

Muscle was segmented by using a threshold of 200 HU outside the ventral cavity after excluding bone and SAT. The muscle mask was post-processed using binary open operation with spherical structuring element of radius 2mm to remove small erroneous labelling.

VAT was segmented by hysteresis thresholding between -200 and 0 HU but inside the ventral cavity. Specifically, two mask volumes were first obtained by thresholding the CT image using -50 HU and 0 HU, respectively, inside the segmented ventral cavity but excluding the segmented lung and voxels with HU < -200 (air pockets in the stomach and bowel). A flood-fill operation was then performed by growing the regions mask volume of -200<HU<-50 into the connected voxels in the mask volume of -200<HU<0. The SAT mask was post-processed using binary open then close operations with spherical structuring element of radius 1mm to remove very small labelling and small holes caused by the small blood vessels in the abdomen and pelvis.

## 3    Results

The ventral cavity segmentation CNN model was trained using the 24 of the 30 datasets from the authors' department and tested on the remaining 6 datasets. The model training and testing process was repeated 5 times by using different remaining 6 testing datasets each time. The overall Dice scores (mean ± standard deviation) for ventral cavity segmentation, measured on the total 30 testing datasets of 5 repetitions were 0.966±0.012. The final CNN model used for body compartment segmentation was trained using all 30 datasets.

The proposed body compartment segmentation procedure was tested on 15 TCGA-KIRC[24] contrast-enhanced CTs and 15 QIN-HEADNECK[12] non-contrast CTs. For each case, bone, VAT, SAT, and muscle were manually contoured on five uniformly sampled slices as the ground truth for quantitative evaluation using three metrics: Dice, recall and precision. The quantitative results are shown in Table 1. Segmentation results for one case are shown in Fig. 5, Fig. 6 and Fig.7.

Table 1. Quantitative evaluations of bone, VAT, SAT, muscle and lung segmentation accuracy for contrast and non-contrast CT.  [a]Spine was segmented as bone.

|  |  | Dice | Recall | Precision |
|---|---|---|---|---|
| Contrast-enhanced CT | Bone[a] | 0.96±0.02 | 0.97±0.03 | 0.95±0.03 |
|  | VAT | 0.94±0.06 | 0.98±0.03 | 0.91±0.09 |
|  | SAT | 0.96±0.04 | 0.98±0.02 | 0.93±0.06 |
|  | Muscle | 0.95±0.04 | 0.95±0.06 | 0.96±0.03 |
|  | Lung | 0.99±0.01 | 0.99±0.01 | 0.99±0.01 |
| Non-Contrast CT | Bone[a] | 0.97±0.02 | 0.97±0.03 | 0.96±0.02 |
|  | VAT | 0.94±0.07 | 0.99±0.01 | 0.89±0.11 |
|  | SAT | 0.93±0.06 | 0.93±0.10 | 0.93±0.04 |
|  | Muscle | 0.91±0.04 | 0.92±0.07 | 0.97±0.02 |
|  | Lung | 0.99±0.01 | 0.99±0.01 | 0.99±0.01 |

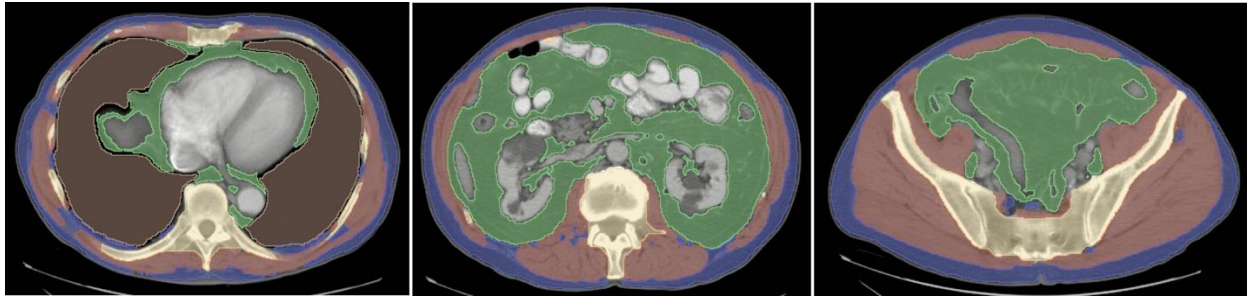

Fig. 5: Segmentation results on selected transverse slices of one contrast CT case. Muscle, SAT, VAT, lung, bone and unsegmented organs are shown in red, blue, green, brown, yellow and gray colors respectively.

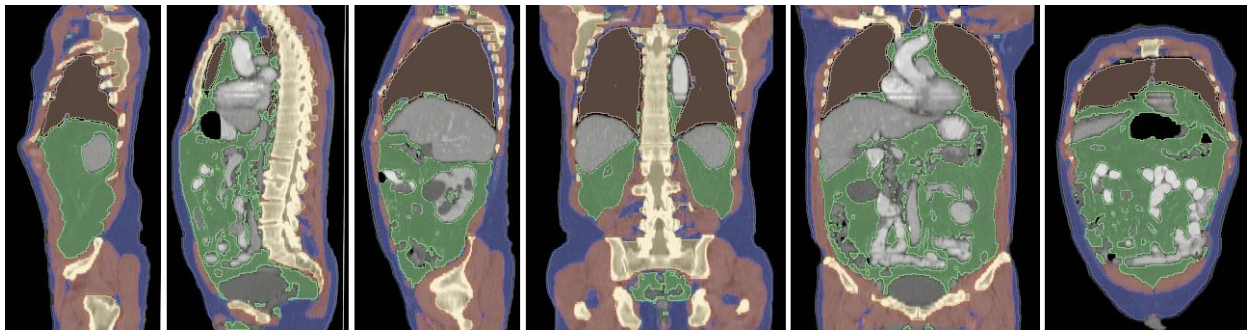

Fig. 6: Segmentation results of selected slices on sagittal and coronal slices of the same case shown in Fig. 5. Muscle, SAT, VAT, lung, bone and unsegmented organs are shown in red, blue, green, brown, yellow and gray colors respectively.

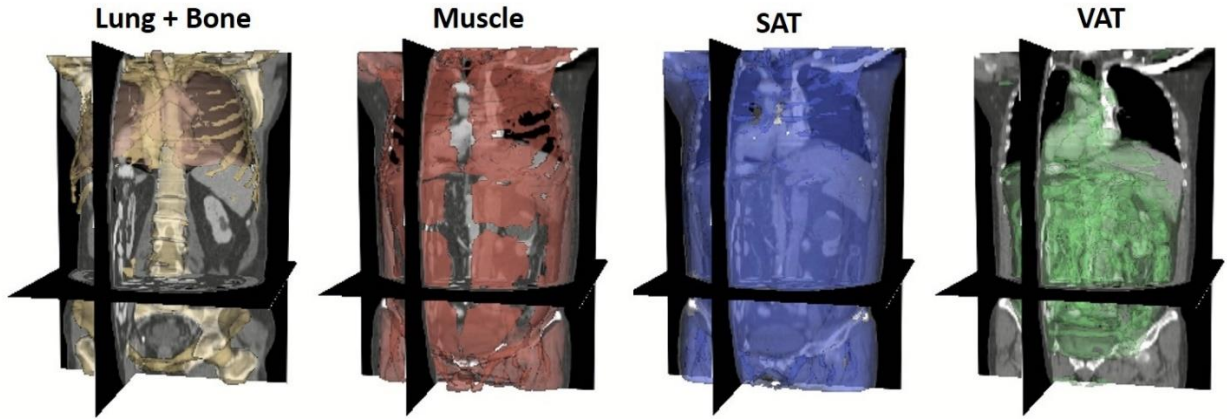

Fig. 7: 3D plots of the segmentation results.

## 4  Discussion

Table 2. Comparison of related works on method, evaluation, region of interest (ROI) and Dice scores.

[a]FCN = convolutional neural network.

| References | | Park et al. (2020)[22] | Weston et al. (2018)[19] | Lee et al. (2017)[13] | Wang et al. (2017)[20] | This study |
|---|---|---|---|---|---|---|
| Methods | | FCN[a] | U-Net | FCN[a] | Slice selection CNN + segmentation CNN | CNN + Postprocessing |
| Evaluating slices (# of testing patients) | | A single axial slice of the third lumbar vertebra (171) | Axial slices of the lumbar vertebras (12) | A single axial slice of the third lumbar vertebra (150) | On average six axial slices across abdomen (20) | Five axial slices across the ventral cavity (30) |
| ROI | | Abdomen | Abdomen | Abdomen | Abdomen | Thoracic + Abdomen + Pelvic |
| Dice | Bone | NA | 0.95±0.05 | NA | NA | 0.96±0.02 |
| | VAT | 0.97 | NA | NA | 0.91±0.06 | 0.94±0.06 |
| | SAT | 0.97 | 0.93±0.06 | NA | 0.98±0.01 | 0.96±0.04 |
| | Muscle | 0.97 | 0.88±0.07 | 0.93±0.02 | NA | 0.95±0.04 |

We compared our method with four related works that have been published in the last three years. Due to different datasets being used, the listed Dice scores only provide a qualitative comparison. Generally, all methods performed well as Dice scores are above 0.90 for most segmented body compartments. Bone usually has the highest Dice score because it is easy to segment. SAT has a higher Dice score than VAT most likely because the presence of fat voxels within stool or gas within the small bowel and colon complicates the VAT segmentation. All methods required laborious manual labeling of many slices for network training. However, we argue that our method requires the least effort because the only structure that requires manual contouring is the ventral cavity, as opposed to other studies that required each individual organs and structures to be manually contoured. One limitation is that 150 2D slices were used to evaluate the body compartment segmentation results. Even though the total number of slices for evaluation is similar to the comparable studies, using significantly more slices for a more thorough evaluation might be necessary in the future work.

In summary, the accurate segmentation results demonstrated a good generalizability of the CNN model. A relatively small number of training datasets were used in this study. It is possible to further improve the performance of the network by including both contrast and non-contrast CTs into our training datasets.

Besides body compartment segmentation, another use case of body cavity segmentation is to model sliding motion in CT abdominal image registration. In non-rigid image registration, spatial smoothing was often used to regularize the calculated displacement vector field. Direction-dependent and anisotropic spatial smoothing has been used to model sliding motion between the lung and chest wall[30,31]. Such spatial smoothing was dependent on tissue classification between sliding organs. Therefore, ventral cavity segmentation can help modelling sliding motion between abdominal wall and internal organs such as lung, liver, and spleen in image registration.

## 5  Conclusion

An image processing procedure was developed to automatically segment seven compartments of the ventral body. The proposed procedure would allow a fully automated computation of 3D ventral body composition metrics in CT images.

## 6  Acknowledgement

The project described was partially supported by the Agency for Healthcare Research and Quality (AHRQ) grant number R01-HS022888, National Institute of Biomedical Imaging and Bioengineering

(NIBIB) grant R03-EB028427 and National Heart, Lung, and Blood Institute (NHLBI) grant R01-HL148210. Its contents are solely the responsibility of the authors and do not necessarily represent the official views of AHRQ, NIBIB or NHLBI.